Article

# Corrosion studies on Fe-30Mn-1C alloy in chloride-containing solutions with view to biomedical application

Annett Gebert, Fabian Kochta, Andrea Voß, S. Oswald, Monica Fernandez-Barcia, Uta Kühn, Julia Hufenbach*

IFW Dresden, Institute for Complex Materials, Helmholtzstraße 20, 01069 Dresden, Germany

Corresponding author: Julia Hufenbach*
IFW Dresden, Institute for Complex Materials, Helmholtzstraße 20, 01069 Dresden, Germany; j.k.hufenbach@ifw-dresden.de

**Abstract**

Austenitic Fe-30Mn-1C (FeMnC) is a prospective biodegradable implant material combining high mechanical integrity with adequate corrosion rates. The fast solidified TWIP alloy, its constituents and 316L stainless steel were electrochemically analysed in various electrolytes at 37 °C under laminar flow. Potentiodynamic polarization tests were conducted in *Tris*-buffered simulated body fluid (SBF), in *Tris*-buffered saline (TBS) and in 150 mM - 0.15 mM NaCl solutions (pH 7.6, 10, 5, 2) to study initial corrosion stages. Active dissolution of FeMnC is revealed in all electrolytes and is discussed on basis of the Fe and Mn behaviour plus is compared to that of 316L. The role of *Tris* (*Tris*(hydroxymethyl)aminomethane) as organic buffer for SBFs is critically assessed, particularly with view to the sensitivity of Fe. SEM studies of FeMnC corroded in NaCl revealed preferential dissolution along Mn-rich grain boundary regions. Static immersion tests of FeMnC in SBF with surface and solution analyses (SEM/EDX, XPS, ICP-OES) indicated that dissolution processes interfere with the formation of permeable surface coatings comprising hydroxides and salts.

Keywords: Fe-based TWIP alloy, biodegradable metal, microstructure, corrosion, simulated body fluid

# 1. Introduction

In recent years, Fe-Mn-based alloys have been developed towards tailored properties for various advanced applications. Austenitic Fe-(10-30)Mn-(Al,Si)-C TRIP/TWIP steels with high strength and ductility as well as excellent formability are of great interest for automotive industry [1] . However, their high Mn content yields only a low corrosion stability in aqueous media, as tested in classical standard solutions like 5-50% $HNO_3$ [2], 5% $Na_2SO_4$, 3.5 -5% NaCl, 0.1M $Na_2SO_4$ or 0.1M NaOH [3, 4, 5]. This has to be surmounted by further alloy modification with passivating elements like Cr [2, 6].

On the other hand, the high reactivity of those Mn-rich alloys may be exploited for medical use, i.e. in the field of biodegradable metal implant devices. Besides Mg-based systems, in particular Fe-based materials are investigated with view to possible osteosynthesis applications [7, 8], as packaging foils for biodegradable electronic devices [9] or as material for stents in the cardiovascular system [10, 11]. Presently, the latter appears to be most promising. Those stents need to be designed for temporarily supporting the tissue healing process while gradually degrading under formation of non-toxic products. Therefore, they must exhibit appropriate mechanical integrity combined with suitable corrosion dissolution within 1-2 years [12, 13]. Pure iron does not meet all these criteria.

Fe-(30-35)Mn austenitic alloys are in the focus of research as they can approach or even exceed mechanical properties of annealed 316L stainless steel which is used as non-degradable stent material. Furthermore, these alloys show antiferromagnetic behaviour, their cytotoxicity (owing to Mn) is still in the allowed limit and they reduce thrombogenicity compared to classical stent materials [11, 14-18]. The high content of Mn in an austenitic Fe-35Mn alloy was found to yield an enhanced corrosion activity in a complex simulated body fluid, e.g. Hank's solution, as compared to pure Fe [19]. The degradation rate can be further enhanced by alloying small amounts of Pd which form Pd-rich precipitates and thus, induce galvanic corrosion phenomena [20]. Fe-Mn-C TWIP alloys with additives like Pd [15] or S [21] are presently of highest interest as they exhibit most promising mechanical performance in terms of high strength and plasticity. Recently, it was demonstrated that by fast solidification alloy microstructures can be obtained which yield this superior mechanical performance without additional mechanical or thermal treatment [21].

With view to biodegradability, all these alloys were tested in established simulated body fluids with complex compositions, i.e. a mixture of different salts with main component sodium chloride and a buffer system. Typical examples are commercial and modified Hank's solution or Dulbecco´s phosphate-buffered saline (DPBS) [19, 22], Dulbecco's modified eagle medium (DMEM) [16] as well as simulated body fluid (SBF) [16] which simulates the inorganic part of human blood plasma. Upon immersion in those solutions, a rather complex surface process occurs comprising metal dissolution and oxygen reduction which is gradually superimposed with hydroxide/oxide formation and precipitation of carbonates, phosphates and other degradation products [23]. Unlike classical metal implants like SS, Co-Cr or Ti alloys, their non-degradability is based on their spontaneous strong passivation ability, these new biodegradable reactive metals are very sensitive to the test conditions. This refers mainly to the fluid composition, i.e. the kind and concentration of ionic species [22] as well as the use of synthetic buffers for pH value control [23]. Further, the fluid flow conditions are decisive, whereby for stent applications dynamic laminar flow is more realistic than static natural convection. Thus, the most suitable test conditions for biodegradable metallic materials are up for current debate. Altogether, despite their prospective application fields Fe-Mn-based alloys are still quite exotic materials in corrosion research and more basic investigations are needed to fundamentally understand their corrosion behaviour.

The present corrosion study on fast solidified Fe-30Mn-1C was conducted with special view to biodegradable metal (stent) applications under laminar fluid flow conditions.

Electrochemical polarization was used to characterize the initial stages of the degradation process. We have stepwise reduced the compositional complexity of the test medium from a Tris-buffered SBF via TBS to a simple NaCl solution in order to assess mainly the effect of the buffer system on the behaviour of a Fe-30Mn-1C alloy. Further, for a more fundamental consideration chloride ion concentration and pH value of NaCl solutions were systematically changed to analyse the sensitivity of the alloy against these fluid conditions. Pure Fe and Mn were used as reference materials to evaluate the impact of the main components and a direct comparison with non-degradable 316L stainless steel was made.

## 2. Experimental

2.1 Materials preparation and characterization

The preparation of the Fe-30Mn-1C (wt.%) alloy was done via a special fast solidification route comprising induction melting of the analytical grade alloying elements in an $Al_2O_3$ crucible up to 1500 °C under argon atmosphere and subsequent ejection into a copper mould [21]. Rod samples with a diameter of 12 mm and a length of 140 mm were obtained. Their chemical composition was checked by means of inductively coupled plasma – optical emission spectroscopy (ICP-OES) (iCAP 6500 Duo View, Thermo Fisher Scientific) and carrier gas hot extraction (Carbon/Sulfur Analyzer Emia – 820V, Horiba Scientific). Microstructural characterization of cast samples was conducted by X-ray diffraction (XRD) (Stadi P, STOE, Mo $K_{\alpha 1}$ radiation). Furthermore, scanning electron microscopy coupled with energy dispersive analysis of X-rays (SEM/EDX) (Leo Gemini 1530, Zeiss/ XFlash4010, Bruker) was applied.

As reference materials analytical grade Fe (>99%) and Mn (>99%) from Womet GmbH as well as 316L stainless steel (Fe-0.03C-(16-18)Cr-(10-14)Ni-(2-3)Mo-2Mn-0.75Si-0.045P-0.03S) provided from Viraj Profiles Ltd. were employed.

## 2.2 Corrosion studies

For electrochemical corrosion tests, the cast Fe-30Mn-1C alloy rods were turned to a diameter of 7 mm, cut into discs with a thickness of 5 mm and embedded in epoxy resin. Similarly, samples of Fe, Mn and 316L were prepared. Both disc faces of each sample were mechanically ground with SiC paper up to grit 4000 and subsequently cleaned with ethanol and dried in air. Those samples served as rotating disc working electrodes. For this purpose, they were attached to an EG&G Parc Model 616B electrode rotor which realizes rotation velocities up to 10.000 rpm. The electronic contact between the backface of the sample and the device was conducted via metal spring. A rotation velocity of 500 rpm was used in order to realize laminar flow conditions of the test medium. The electrochemical cell comprised also a Pt net as counter electrode and a SCE reference electrode (E(SHE)=0.241 V) and was connected to a Solartron SI 1287 electrochemical interface. All measurements were done in a tempered double-wall glass cell at controlled 37±1°C. The following test solutions were used:

i) SBF (Simulated Body Fluid) prepared according to Kokubo et al. [24] containing 142 mM $Na^+$, 5 mM $K^+$, 1.5 mM $Mg^{2+}$, 2.5 mM $Ca^{2+}$, 147.8 mM $Cl^-$, 4.2 mM $HCO_3^-$, 1.0 mM $HPO_4^-$, 0.5 mM $SO_4^{2-}$; the pH 7.45-7.6 was buffered with 50 mM *Tris* (*Tris*(hydroxymethyl)aminomethane , $(HOCH_2)_3CNH_2$)) and 1 M HCl.

ii) TBS (*Tris*-Buffered Saline, Fa. Sigma Aldrich) a 150 mM (~0.9%) NaCl solution buffered with 50 mM *Tris* to pH 7.6.

iii) NaCl - non-buffered solutions made from high purity NaCl (Fa. Merck); in a *first series* solutions with 150 mM NaCl were prepared and initial pH values of 7.6, 5, 2 and 10 were adjusted with HCl and NaOH, respectively; in a *second series* solutions with 150 mM, 15 mM, 1.5 mM and 0.15 mM NaCl were prepared and a pH value of 7.6 (pH 10) was adjusted with NaOH.

After recording the open circuit potential (OCP) for 1 h under rotational flow conditions, potentiodynamic polarization measurements were conducted starting from -0.15 V vs. OCP up to an end potential of 1.5 V vs. SCE using a potential scan-rate of 0.5 mV/s. All measurements were repeated at least 3 times to evaluate the reliability of the data. From the polarization curves the corrosion parameters $E_{corr}$ and $i_{corr}$ were derived by means of the graphical Tafel extrapolation method. After corrosion testing under free corrosion conditions (OCP, 4 hours) and after anodic polarization tests selected alloy samples were subjected to SEM investigations in order to derive information regarding the initial corrosion steps in NaCl-based solution.

For analyzing the degradation mechanism of Fe-30Mn-1C in SBF solution, static immersion tests were performed in 450 ml SBF at 37 °C for different durations (24 h, 72 h and 168 h) with each two samples. Therefore, the specimens with a thickness of 1.5 mm and diameter of 12 mm were cut out of the cast rod and ground up to 4000 SiC grit paper.

By using ICP-OES, metal release of Mn and Fe was determined after immersion tests (filtrate=dissolved fraction). Additionally, the solid corrosion products were analysed by ICP-OES by solving the products in a hot solution containing 5 ml hydrochloric acid (HCl), 250 mg oxalic acid ($C_2H_2O_4$) and 5 ml distilled water (precipitate=solid fraction). For each tested sample, a four-fold determination was conducted to determine the average element concentration of Fe and Mn. Thereby the methodological error was below 1 %.

For the analysis of surface coatings on alloy samples, sputter depth-profiling X-ray photoelectron spectroscopy (XPS; Physical Electronics, PHI 5600 CI, Mg Kα radiation, 400 W) was applied. Concentration quantification was done using single element standard sensitivity factors. For XPS analysis, samples were ground up to 4000 SiC grit paper and polished with 0.25 µm diamond suspension (Buehler), whereby one sample was exposed to SBF for 1h (under the same conditions like for the immersion tests). For sputtering $Ar^+$ ions (3.5 keV) were used removing a surface layer with a rate of about 3.3 nm/min measured on $SiO_2$.

## 3. Results and Discussion

3.1 Microstructural characterization of Fe-30Mn-1C

Characterization of the fast solidified Fe-30Mn-1C alloy by means of XRD (not shown here) revealed an austenitic phase with *Fm-3m* structure. Fig. 1 summarizes a typical SEM image taken from the cross-sectional area of a rod sample and the corresponding EDX mappings of Fe, Mn and C. A dendritic microstructure is presented which is composed of fine austenitic grains [21]. From the EDX analysis an enrichment of Mn along the austenite grain boundaries can be derived whereby the core regions of the austenite grains are Fe-rich relative to the nominal alloy composition. Carbon appears to be nearly homogeneously distributed. The beneficial mechanical performance data have been described in a previous study [21].

3.2 Comparative electrochemical corrosion studies in SBF, TBS and NaCl

In order to evaluate the initial corrosive steps of the alloy degradation process in simulated body fluid, electrochemical polarization studies were conducted under laminar electrolyte flow conditions. Fig. 2a shows typical current density-potential curves for the austenitic Fe-30Mn-1C alloy as well as of its main constituents Fe and Mn and of the 316L reference steel recorded in SBF (pH 7.45-7.6) after exposure for 1 h under open circuit conditions. To assess the influence of the compositional complexity and ionic strength of this medium as well as of the *Tris*-buffer also polarization curves were measured in simplified media. These are TBS (*Tris*-buffered 150 mM NaCl, pH 7.6) and non-buffered 15 mM NaCl (pH 7.6). The typical polarization curves are presented in Fig. 2b and c for comparison.

In SBF, a key feature is the abnormality in the curve of Fe. It exhibits in the low polarization regime several sharp current density minima before it transfers into an active dissolution regime at higher anodic potentials. Such a behaviour is similarly evident in TBS, whereby in the simple NaCl solution the classical behaviour with one sharp current density minimum is observed followed by an immediate active dissolution under anodic conditions. Therefore, this abnormality may be attributed to an effect of the organic buffer *Tris* on the surface reactions of Fe. In earlier papers, the initial Fe reactions in Hank's solution [25] or SBF [17] were simply described with the electrochemical oxidation to $Fe^{2+}$ and further competitive formation of $Fe(OH)_2$/$Fe(OH)_3$ or direct reaction with $Cl^-$ to $FeCl_2$ which hydrolyses to $Fe(OH)_2$. However, according to MacFarlane and Smedley [26] the dissolution mechanism of Fe in chloride solutions at low polarization conditions is more complex. It comprises the electrochemical formation of $[FeClOH^-]_{ads}$ adsorbate and its oxidation to FeClOH (rate-determining step) which further chemically decomposes under formation of $Fe^{2+}$. In presence of oxygen this can further gradually oxidize to $Fe^{3+}$.

*Tris* is the short term for *Tris*(hydroxymethyl)aminomethane which tends to from Fe(II)- or Fe(III)-complexes with relative stability already at low ion concentrations [27]. Therefore, in *Tris*-buffered chloride-containing solutions the Fe dissolution may be distorted by this complex formation. The appearance of several minima in the polarization curve hints to temporary surface dissolution related with current density rise and subsequent short surface blocking (minima) due to a covering of the Fe surface with low or no conducting adsorbates. Thus, the Fe dissolution is inhibited over a wider potential range. Only at higher anodic potentials (where the role of chloride ions is less important [26]) the expected severe active dissolution behaviour is observed.

The polarization curves of the other main alloy constituent Mn do not comprise such an abnormal behaviour. In all three test solutions Mn is very reactive and dissolves actively, as evident from very negative corrosion potentials $E_{corr} \ll 1$ V vs. SCE, which are about 570-670 mV more negative than values for Fe. Further, the corresponding corrosion current densities $i_{corr}$ of 0.04 – 0.6 mA/cm² are very high and a steep increase of the anodic current density up to the mass-transfer controlled regime occurs. In the SBF and TBS, corrosion current densities for Mn are about one order of magnitude higher than those for Fe, but in non-buffered NaCl values they are on a similar level.

When solving Mn and C in austenite (γ-Fe), the above discussed abnormalities in the low polarization regime of the current density-potential curves measured in *Tris*-buffered solutions do not occur. In all three test electrolytes, for the Fe-30Mn-1C alloy the corrosion potentials establish in between those for Fe and Mn, but in closer vicinity to that of the main constituent Fe. Furthermore, the corrosion current densities are in the typical range of that for Fe. Upon anodic polarization the immediate strong rise of the current density indicates a strong alloy dissolution up to the mass transfer-controlled regime. The non-appearance of the distortions of the alloy dissolution in *Tris*-buffered solutions may be attributed to the demonstrated strong reactive nature of Mn which can counteract the above discussed effect of possible Fe-complex formation. However, in an earlier study it was shown that also the addition of C to Fe (Fe-1C) can induce such an effect [21].

Altogether, the electrochemical response of such an austenitic Fe-30Mn-1C alloy and thus, the initial corrosion stages appear to be much less sensitive to effects of the *Tris* buffer than that of pure Fe. Nonetheless, with view to these discussed phenomena the use of synthetic organic buffers like *Tris* (which are commonly employed for in vitro studies in biochemistry but are not present in the human body) for analysing the degradation behaviour of the new biodegradable metallic materials should be seen critical. This conclusion is in line with those of Schinhammer et al. [17] or Mouzou et al. [22] who deduced from long-term exposure studies with Fe-Mn-C alloys that the complex reaction scheme of the alloy degradation in a simulated body fluid can be significantly modified by those organic compounds. Similar discussions are currently raised for Mg-based materials [28, 29] and underline the necessity for a new definition of standard test conditions for those new materials for medical applications.

A general comparison of the polarization curves measured in the three test electrolytes reveals that in the low polarization regime the austenitic Fe-30Mn-1C alloy and its main constituents have the lowest reactivity in SBF. Besides buffer effects, this may be attributed to the complex solution composition with anions like $HCO_3^-$ and $HPO_4^-$ that tend to salt formation and thus, may inhibit the chloride-driven metal dissolution.

Moreover, to emphasize the reactive nature of the austenitic alloy, comparative measurements on 316L steel were conducted (Fig. 2a-c). Typically, for 316L the corrosion potential values are the most positive ones and corresponding corrosion current density values are 1 to 3 orders of magnitude lower than those of the biodegradable alloy. This indicates a very low reactivity and can be attributed to the high Cr and Ni contents which yield spontaneous passivation of the steel surface and thus, inhibit dissolution reactions. Upon anodic polarization a passive plateau is reached before pitting is initiated corresponding to a steep rise of the current density.

3.3 Corrosion analysis of Fe-30Mn-1C in SBF

In order to characterize the surface reactions of the austenitic Fe-30Mn-1C alloy in the complex SBF solution in more detail, also immersion tests up to 7 days were conducted. While the polarisation studies presented in chapter 3.2 reflect only the electrochemical formation of constituent ions $Fe^{2+}$ and $Mn^{2+}$ and corresponding dissolved oxygen reduction, adequate surface and solution analyses related with immersion tests can deliver deeper inside into the complex reaction process.

Fig. 3 shows SEM images of characteristic surface states of the alloy after selected immersion durations in SBF. After 1 hour the principal features of the dendritic microstructure (compare

Fig. 1) are still recognizable. But also numerous small holes are present which indicate pitting, preferred either in Mn-rich grain boundary regions or at MnS inclusions. After further immersion up to 4 hours the base features vanished which hints to enhanced surface coverage and local defects (pits) are more developed and covered with precipitates. These processes further evolve and lead after 3 days of immersion to multiple larger pits which are partially filled with precipitates. Further, the surface coating on the surrounding matrix appears to be more coarse-grained. A larger defect with the surrounding surface area which was identified after 7 days of immersion is shown in Fig. 3d. The defect has a flat morphology with an inner corroded region, whereby the rim region has a crater-like shape revealing a local peeling off of the otherwise homogeneous and adhesive surface coating as well as additional local corrosion product precipitation.

Sensitive XPS analysis has been applied to characterize the initial surface stages of the Fe-30Mn-1C alloy after short-term immersion in SBF relative to an as-polished reference state. Typical XPS depth profiles are shown in Fig. 4. Deviations in the measured atomic concentrations of the alloy constituents can result from both the use of single-element sensitivity factors and from surface de-mixing by preferential sputtering at the multicomponent material. Already in the as-polished state the alloy is covered with a natural surface film. At the first sputter steps besides Fe and Mn also O species are present in the depth profile indicating a spontaneously forming passive film of Fe- and Mn- oxides with a thickness of a few nanometres (Fig. 4a). However, after immersion in SBF for only 1 hour a significant broadening of the oxygen (O) depth profile is detectable (Fig. 4b) which reveals the presence of a surface coating with a thickness of 40-50 nm (considering the sputter time at 50 % decrease of the oxygen signal and the standard sputter rate). In near neutral aqueous solutions mainly the hydroxides $Fe(OH)_2/Fe(OH)_3$ and $Mn(OH)_2$ are stabilized which can form by direct reaction of the dissolved ionic metal species with generated hydroxide ions or as result of the hydrolysis of metal-chloro complexes. In addition, Ca and P species are detected in the surface region which suggest precipitation of phosphate compounds already at this early state of immersion. Also C species are detectable which may be attributed to contaminations of the outermost surface region but also to the formation of carbonate compounds [12,17].

Fig. 5 shows results of EDX mapping of an area with a larger defect and the surrounding surface region formed after several days of immersion in SBF (compare Fig. 3d). The initial thin layer grew with prolonged immersion time significantly and peeled off at surface spots where pitting occurred which is related with a locally enhanced metal degradation. The corrosion front propagates predominantly in lateral direction and causes local stress-induced cracking of the coating. From the elemental maps it can be derived that not only O-rich compounds are present in the defect zone, but also Ca- and P-species are accumulated as well as Mg- and Cl-species. This confirms that besides hydroxides also progressing formation of Ca-phosphates (tricalcium phosphate or hydroxyapatite) occurred together with precipitations from SBF like $MgCl_2$.

In addition to those surface analytical studies, also chemical analysis of the immersion solution with respect to Fe and Mn species was conducted. Already after 1 day of immersion, the test medium comprised besides the solution fraction with dissolved metal ion species also a solid fraction which is attributed to both, precipitates forming in the solution due to local oversaturation in the static medium and detachments from the layer which grows on a sample surface. Fig. 6 summarizes results of ICP-OES analyses after 1, 3 and 7 days of immersion. It is obvious from Fig. 6a, that the SBF solutions contain only $Mn^{2+}$ species with increasing concentrations, whereby $Fe^{2+}$ concentrations remain below the quantification limit after all immersion durations. On the contrary Fig. 6b shows, that the analysis of the solid fractions revealed mostly Fe species and only significantly smaller concentrations of Mn species which marginally increase with prolonged immersion time. According to these analysis results, Mn demonstrates in SBF solution under static conditions a high tendency to be present in ionic state. Dissolved $Mn^{2+}$ ions can be stabilized in concentrated chloride-containing solutions as metal-chloro complexes (e.g. $MnCl_3^-$) and only a smaller fraction of solid compounds with low solubility, e.g. $Mn(OH)_2$ or $MnCO_3$, precipitate. On the contrary, the predominant fraction of $Fe^{2+}$ transforms into solid products with low solubility, i.e. mostly $Fe(OH)_2$/$Fe(OH)_3$ and $FeCO_3$ [12,17,22].

Altogether, these immersion tests confirm that in the complex simulated body fluid SBF the active corrosive dissolution of the Fe-30Mn-1C alloy is superimposed by the formation of surface coatings comprising hydroxides and salts (carbonates) of the constituents as well as precipitates from the concentrated solution (Ca-phosphates, Mg-chlorides). These coatings can to a certain degree retard the degradation process, but they are still permeable enough to enable metal release even after several days. This is eased by local coating breakdown due to pitting events. However, these findings are derived from immersion tests under static conditions. Future studies will refer to long-term immersion analysis under laminar flow control to better simulate the stents application situation.

3.4 Corrosion studies in non-buffered NaCl solutions

In another part of the study the corrosion behaviour of Fe-30Mn-1C and its constituents was looked at in a more fundamental manner. The sensitivity of this new alloy to changes of the pH value and of the chloride concentration of the non-buffered flowing electrolyte was evaluated. In a first series, starting from the 150 mM NaCl solution with pH 7.6 (Fig. 2c) the pH value was changed. Fig. 7 shows selected current density – potential curves recorded in solutions with pH 2 and pH 10. The electrochemical parameters $E_{corr}$ and $i_{corr}$ were determined form those curves and are summarized in Fig. 9 together with data taken from the previous measurements (curves in Fig. 2a-c). At both, more acidic and more alkaline conditions the actively dissolving nature of the Fe-30Mn-1C alloy and its main constituents Fe and Mn is preserved. This suggests that the high chloride ion concentration in the electrolytes governs the surface reactions. Therefore, $E_{corr}$ values change only marginally. However, with reduction of the pH value to pH 5 and pH 2, $i_{corr}$ values increase remarkably, i.e. by up to one order of magnitude. This demonstrates that the acidification of the solution enhances the corrosion rates at low polarization. Consequently, the alkalization of the chloride-containing solution reduces the corrosion rates. On the contrary, for the 316L steel the corrosion current density remains low, i.e. <1 µA/cm², even at pH 2. This is attributed to its strong spontaneously passivating nature. However, with shift of the pH value from pH 10 to pH 2 the anodic potential region of the passive plateau diminishes and pitting occurs earlier. This indicates that the permeability of growing oxide films increases. Altogether, at all pH value levels of the 150 mM NaCl solution, the significant difference in the electrochemical responses of the new Fe-30Mn-1C alloy and the established 316L steel is evident.

In a next step, the effect of chloride ion concentration in solutions with pH 7.6 was assessed. Therefore, polarization measurements were conducted in 150 mM, 15 mM, 1.5 mM and 0.15 mM NaCl solutions. Exemplarily, the current density – potential curves recorded in the most diluted solution 0.15 mM NaCl are shown in Fig. 8. Corrosion parameters $E_{corr}$ and $i_{corr}$ are plotted in dependence of the NaCl concentration in Fig. 9. Even at the lowest chloride ion concentration tested, the austenitic Fe-30Mn-1C alloy remains in an actively dissolving state. But with decreasing salt concentration the corrosion potential $E_{corr}$ shifts to more positive values and the corrosion current density decreases visibly up to two orders of magnitude. This indicates a significant diminution of the corrosion activity of this alloy. For Mn a similar trend in the behaviour was observed. Fe is most sensitive to the salt concentration change. Its corrosion activity also reduces with decreasing chloride ion concentration. In low concentrated solutions, i.e. 1.5 mM and 0.15 mM NaCl, it exhibits limited passivity under anodic conditions. Thus, its polarization behaviour becomes more similar to that of the 316L steel. A certain passive potential range is followed by pit initiation with gradual current density rise indicating slow pitting rates. In additional measurements conducted in the most diluted salt solution 0.15 mM NaCl but with higher pH value of 10 (not shown here, corrosion data in Fig. 9) the principal electrochemical responses of the tested materials did not significantly change. Obviously, in the frame of the tested corrosion conditions, the high Mn content of the new austenitic Fe-30Mn-1C alloy suppresses surface passivation that could yield a protective effect on chloride ion-containing media.

The retarded corrosion rates in diluted NaCl solution allow observations of the initial corrosion stages of the new austenitic alloy. For that Fe-30Mn-1C samples were mildly corroded in 0.15 mM NaCl solution with pH 7.6 and the sample surface were subsequently analysed with SEM. Fig. 10 shows typical corroded surface morphologies after exposure under open circuit conditions for 4 hours and after a polarization measurement. Under OCP conditions (Fig. 10a and b), the alloy surface is partly attacked. A sharp corrosion front exist between smooth un-corroded regions and corroded regions in which the pattern of the microstructure is recognizable. After anodic polarization (Fig. 10c and d) the surface dissolution is more progressed and the grain structure is more clearly revealed. These general features indicate a stronger dissolution of the Mn-rich austenite grain boundaries. This corresponds to the discussed results of electrochemical corrosion analysis, i.e. that Mn enhances the dissolution activity of the austenitic phase. Moreover, in particular in the higher resolved SEM images sub-structures are recognizable in the austenite grains which can be led back to preferential corrosion attack depending on grain orientations and microsegregations. Those features are already known for other corroding metals and have been described in detail, for example, for Mg [30]. Further, Fig. 10c and d show also a few single spherical pits (indicated by arrows), some of them contain round precipitates. The fast solidified Fe-30Mn-1C alloy contains a few small MnS inclusions formed due to the existence of S traces (accompanying element of Mn) which are well known as spots for pitting initiation [16]. Altogether, these microscopic observations of initial corrosion states reveal a non-uniform surface degradation of the new austenitic alloy. In future studies, its impact on the mechanical performance will be evaluated.

## 4. Summary and conclusion

In electrochemical polarization studies und laminar flow conditions, the austenitic Fe-30Mn-1C alloy exhibited in all test solutions – from complex *Tris*-buffered SBF to non-buffered NaCl solutions with either very low concentration or strong acidification – an actively dissolving nature. This is very different to the behaviour of 316L stainless steel, its low corrosion rates are due to a strong passivation tendency. An activity enhancement of the austenitic alloy relative to Fe, due to the presence of reactive Mn, is mainly evident from a general shift of the corrosion potential to more negative values. In *Tris*-buffered SBF, for the ternary alloy the lowest reactivity was observed. Additional immersion tests demonstrated that in in this complex solution chloride-induced dissolution processes are retarded already at an early stage by hydroxide formation and salt precipitations.

In low concentrated NaCl solutions an effect of Mn in yielding significantly higher corrosion current densities of Fe-30Mn-1C compared to passivating Fe is detectable. Moreover, local Mn enrichment in austenite grain boundaries causes non-homogeneous surface dissolution in the initial corrosion stage and thus, formation of corrosion patterns. The role of *Tris* as organic buffer for simulated body fluids is critically assessed in particular with view to the sensitivity of the electrochemical response of Fe. This study supports the need for definition of suitable test conditions for the new class of biodegradable metals.

Beyond those fundamental studies for assessing principal properties of the new alloy, future work will focus on application-relevant questions. Thereby, the impact of different processing routes for stent production on the microstructure of the alloy and the consequences for mechanical, corrosion, biological and magnetic properties will be analysed. This includes for example the application of additive manufacturing technologies like selective laser melting as well as the development of suitable electropolishing procedures for those reactive biodegradable materials.

*Acknowledgements*

The authors thank M. Frey, M. Johne and R. Keller for technical support. Funding of this work by the German Research Foundation (DFG) under project number HU 2371/1-1 and by the European Commission with the H2020-MSCA-ITN-2014 SELECTA, grant agreement no. 642642 is gratefully acknowledged.

**Figure captions**

**Fig. 1**  (a) SEM image and corresponding EDX mappings of (b) Fe, (c) Mn and (d) C revealing the microstructure of a fast solidified Fe-30Mn-1C alloy sample.

**Fig. 2**  Current density-potential curves of austenitic Fe-30Mn-1C, Fe, Mn and 316L SS recorded in (a) SBF (pH 7.45-7.6), (b) TBS (pH 7.6) and (c) 150 mM NaCl (pH 7.6) (scan rate 0.5 mV/s; RDE 500 rpm).

**Fig. 3**  SEM images of Fe-30Mn-1C surfaces after static immersion in SBF at 37 °C: (a) 1 h, (b) 4h, (c) 3 days and (d) 7 days.

**Fig. 4**  XPS depth profiles of Fe-30Mn-1C surfaces: (a) as-polished and (b) after static immersion for 1 h in SBF at 37 °C (total concentrations of the measured elements).

**Fig. 5**  (a) SEM image of a corroded area of Fe-30Mn-1C after 7 days static immersion in SBF and corresponding EDX mappings of (b) Ca, (c) P, (d) O, (e) Mg and (f) Cl.

**Fig. 6**  ICP-OES analysis data**:** (a) dissolved and (b) solid fractions of released Mn and Fe species in SBF after 1 day, 3 days and 7 days of static immersion (Fe-30Mn-1C alloy).

**Fig. 7**  Current density-potential curves of austenitic Fe-30Mn-1C, Fe, Mn and 316L SS recorded in 150 mM NaCl solution with (a) pH 2 and (b) pH 10

(scan rate 0.5 mV/s; RDE 500 rpm).

**Fig. 8** Current density-potential curves of austenitic Fe-30Mn-1C, Fe, Mn and 316L SS recorded in 0.15 mM NaCl solution with pH 7.6
(scan rate 0.5 mV/s; RDE 500 rpm).

**Fig. 9** Corrosion parameters (a) $E_{corr}$ and (b) $i_{corr}$ determined from current density-potential curves measured in the different test electrolytes; NaCl-variation of concentration (at pH 7.6) and variation of pH value (at 150 mM NaCl).

**Fig. 10** SEM images of corroded surfaces of Fe-30Mn-1C samples after (a and b) exposure for 4 h under open circuit conditions and (c and d) current density-potential curve measurement in 0.15 mM NaCl (pH 7.6).

**Graphical Abstract**

**Table of Contents**

The Fe-30Mn-1C TWIP alloy is a prospective material for biodegradable implants with tailored microstructures yielding suitable properties. This work looks at the initial stages of the alloy degradation. Its actively dissolving nature is shown in different electrolytes from SBF to simple NaCl solutions and discussed based on Fe and Mn behaviour and relative to that of 316L SS. The use of Tris-buffered solution for evaluating biodegradability is critically assessed.

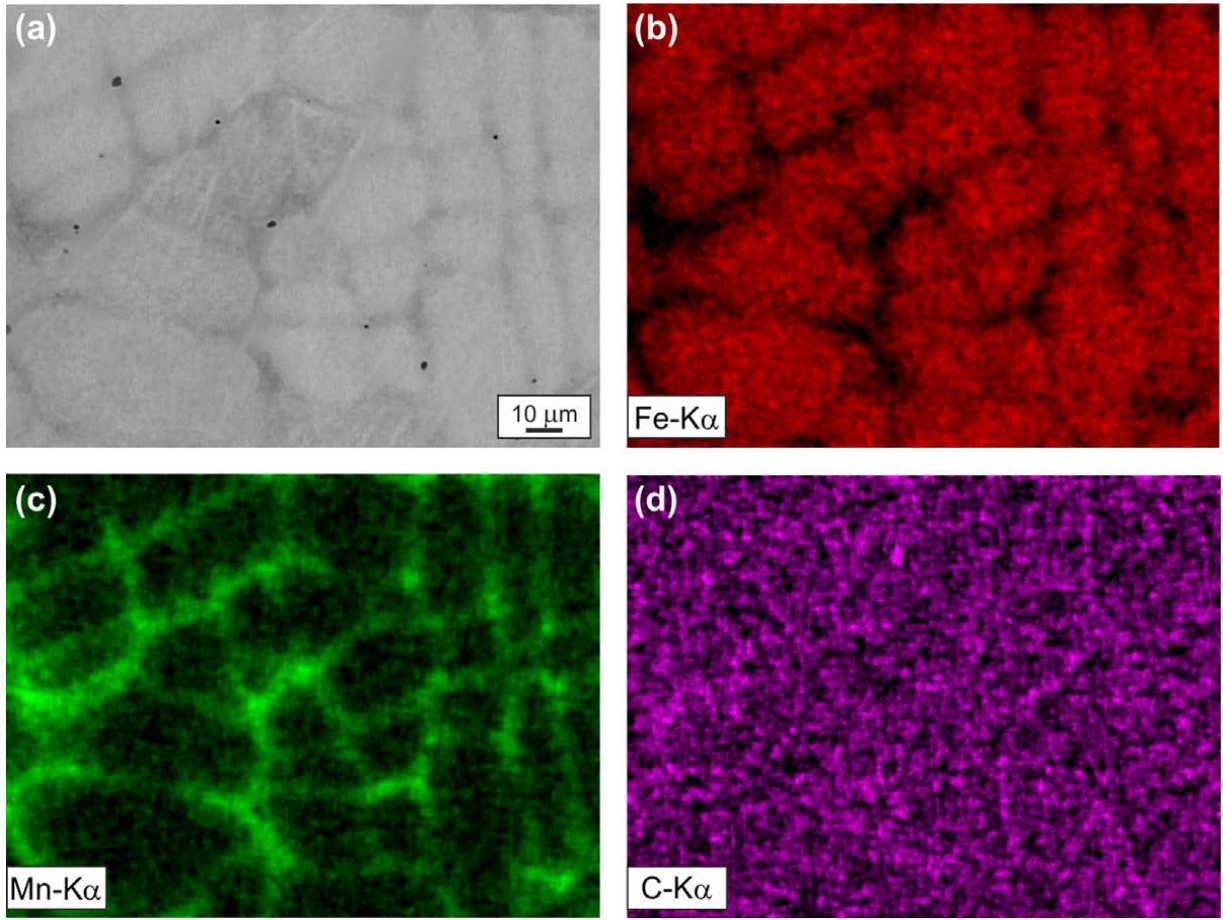

Figure 1

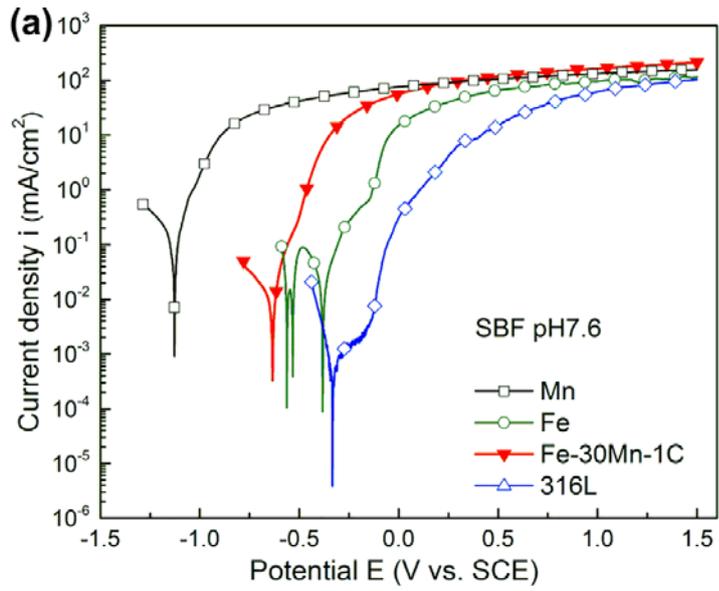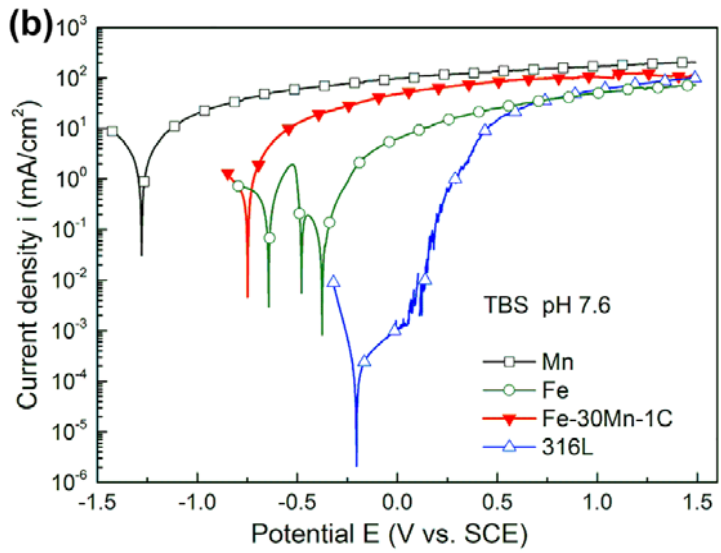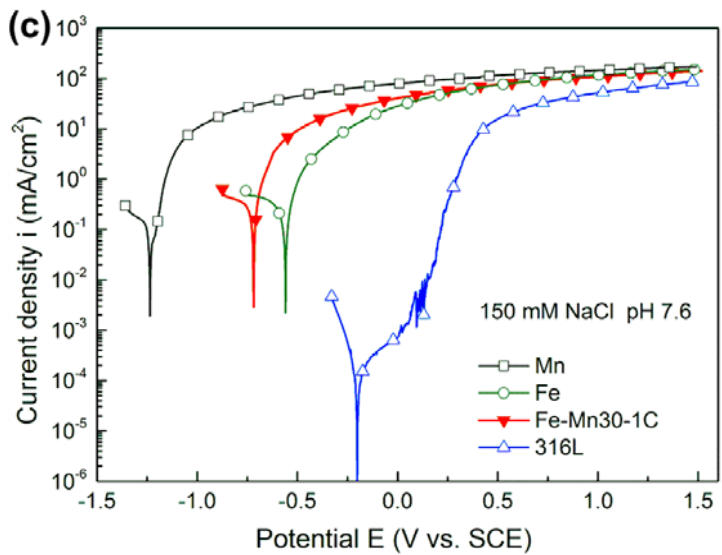

Figure 2

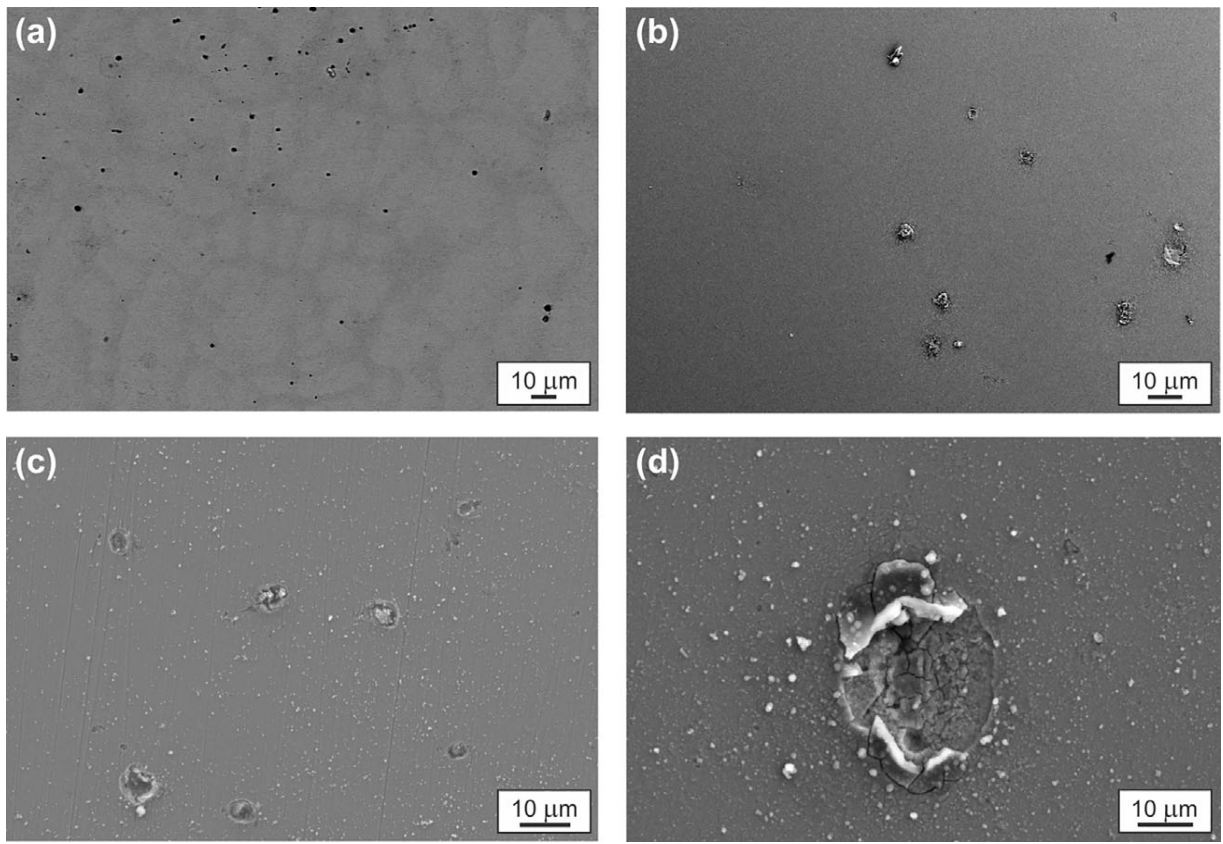

Figure 3

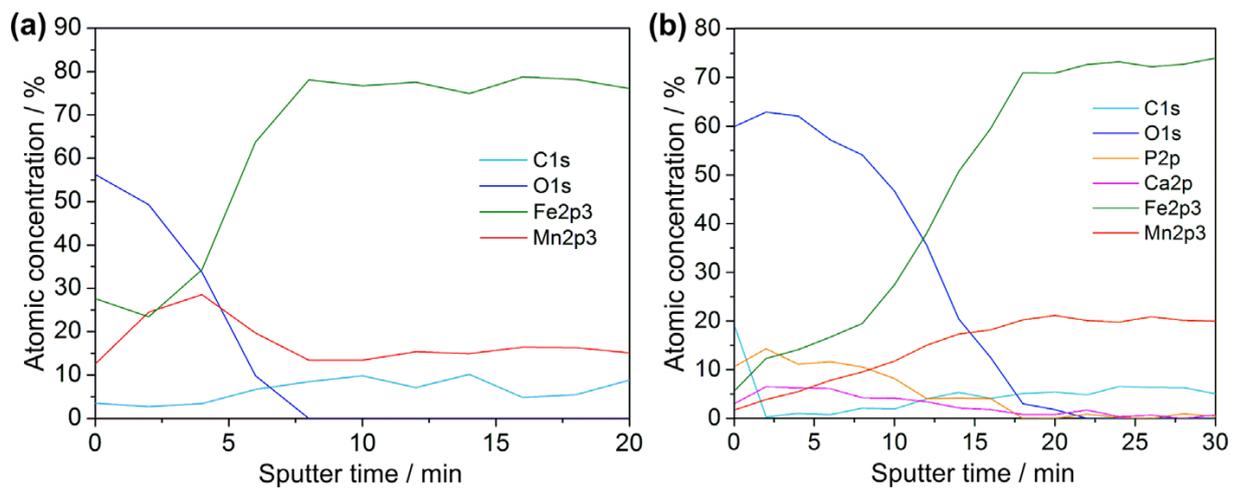

Figure 4

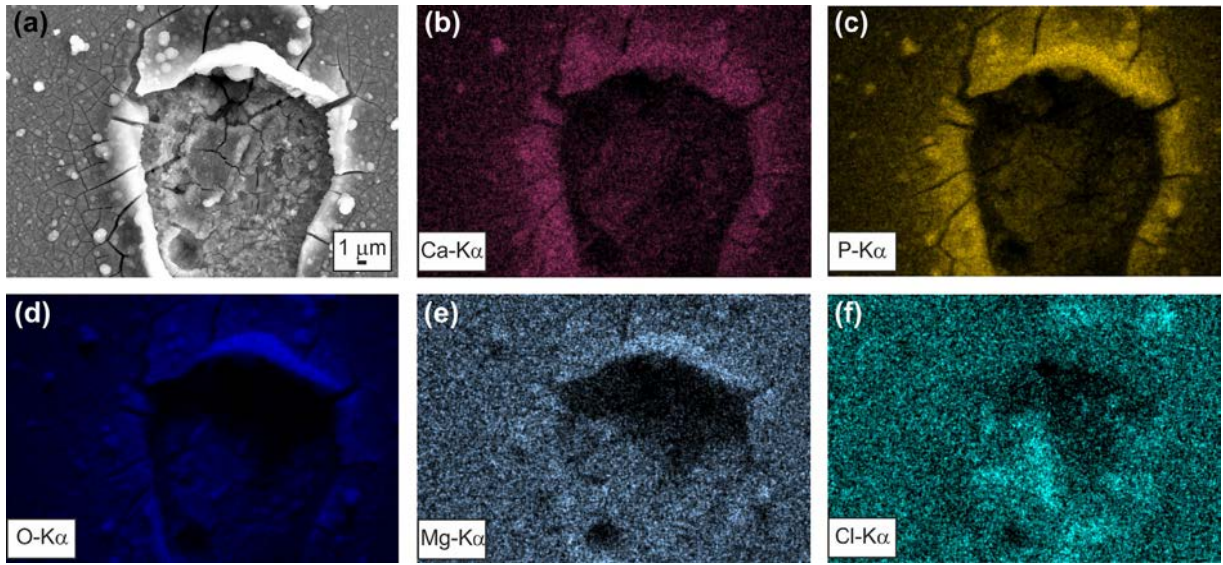

Figure 5

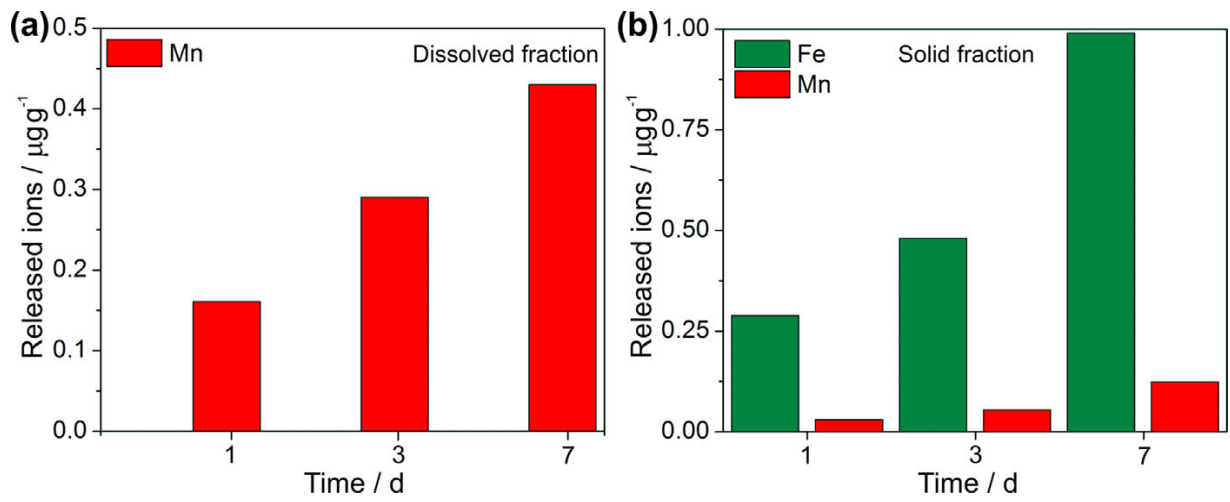

Figure 6

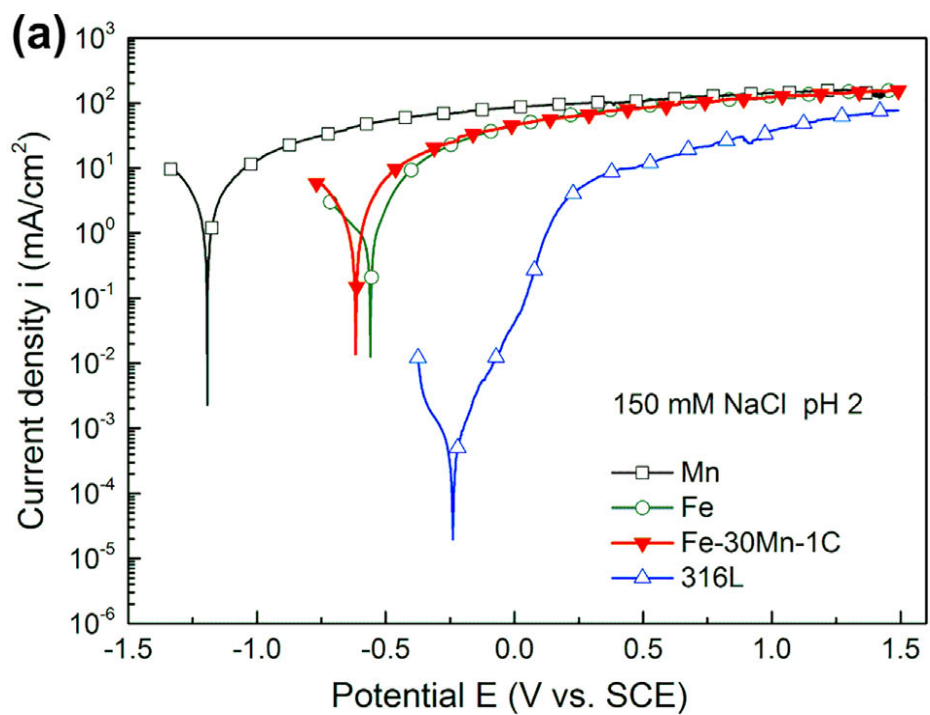

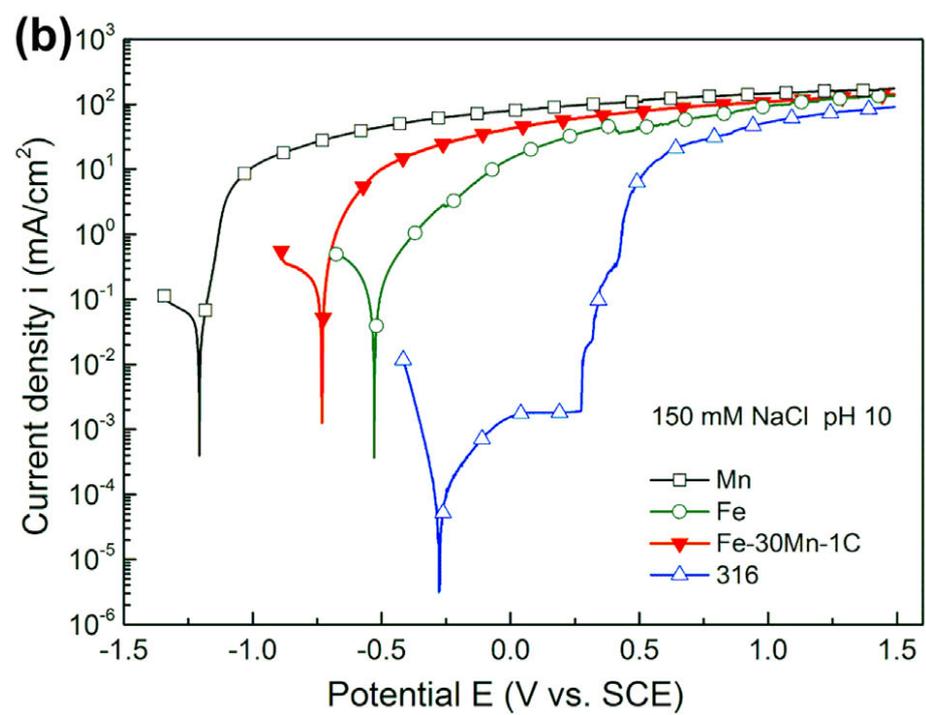

Figure 7

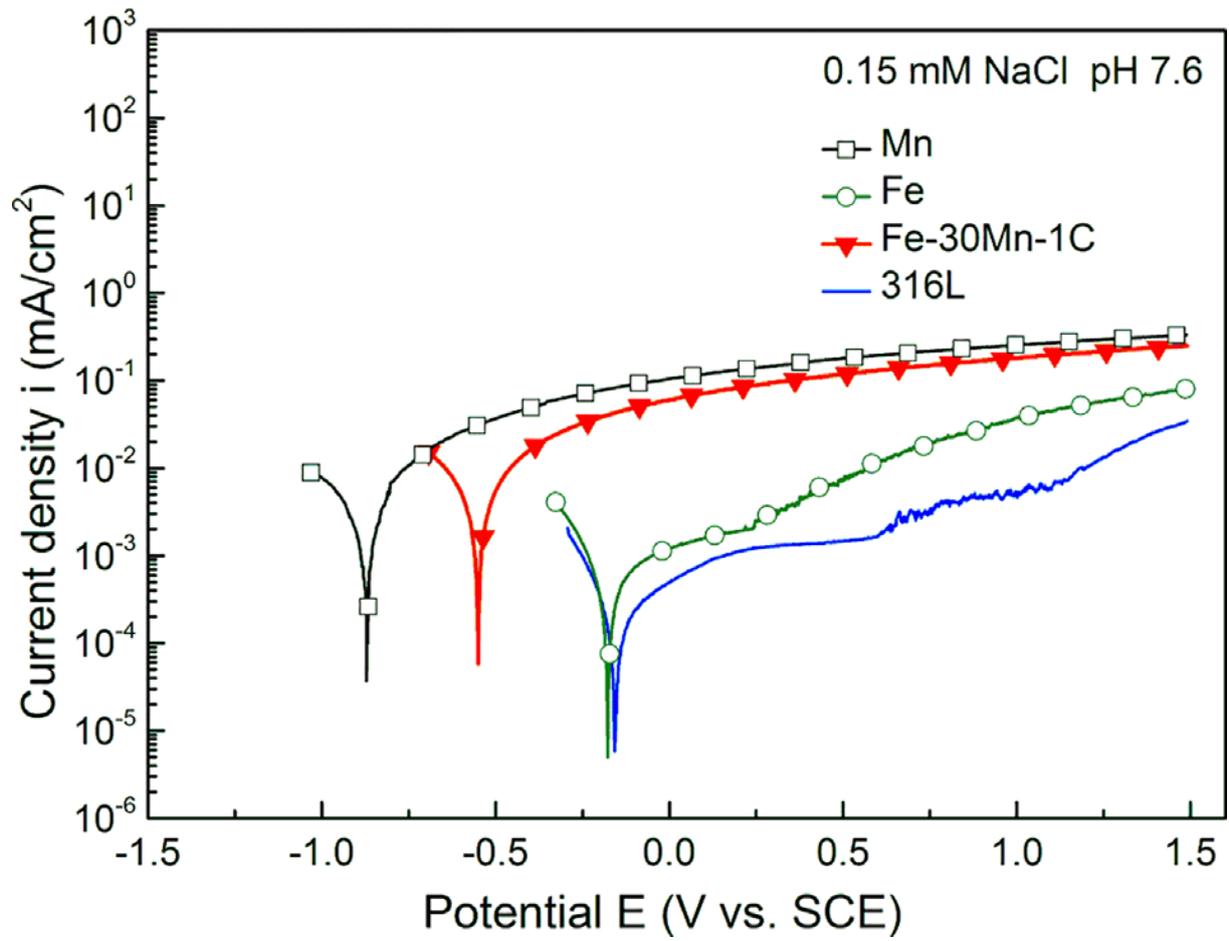

Figure 8

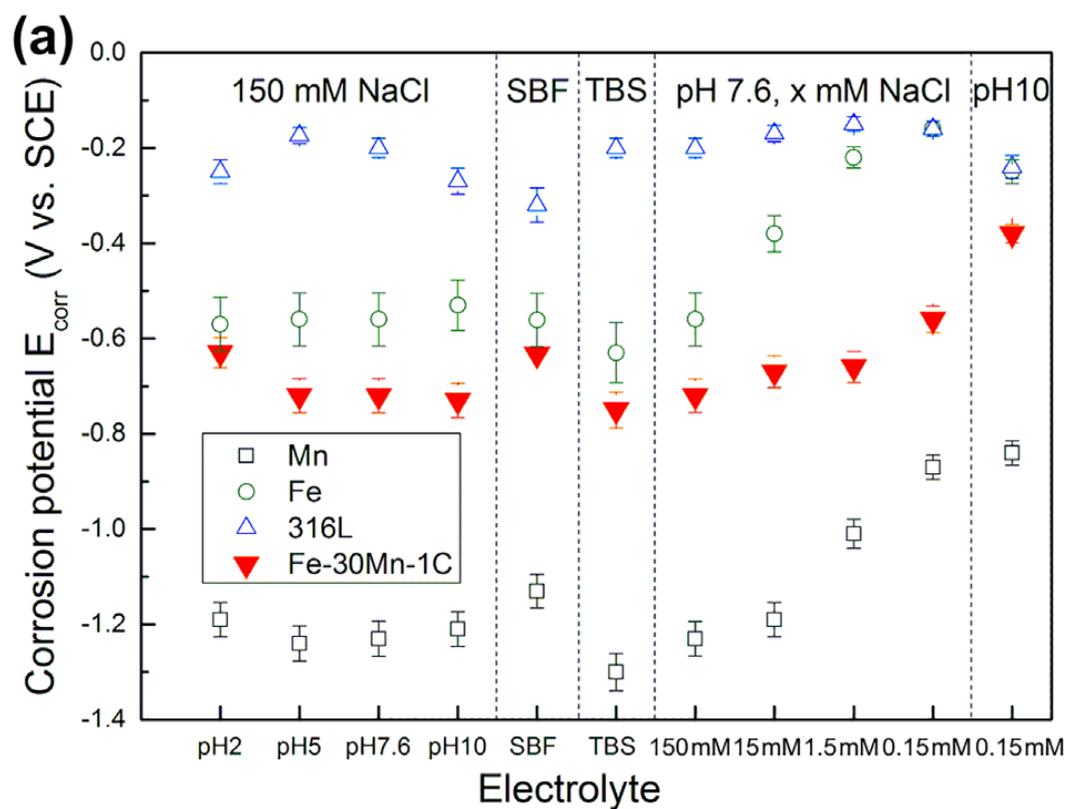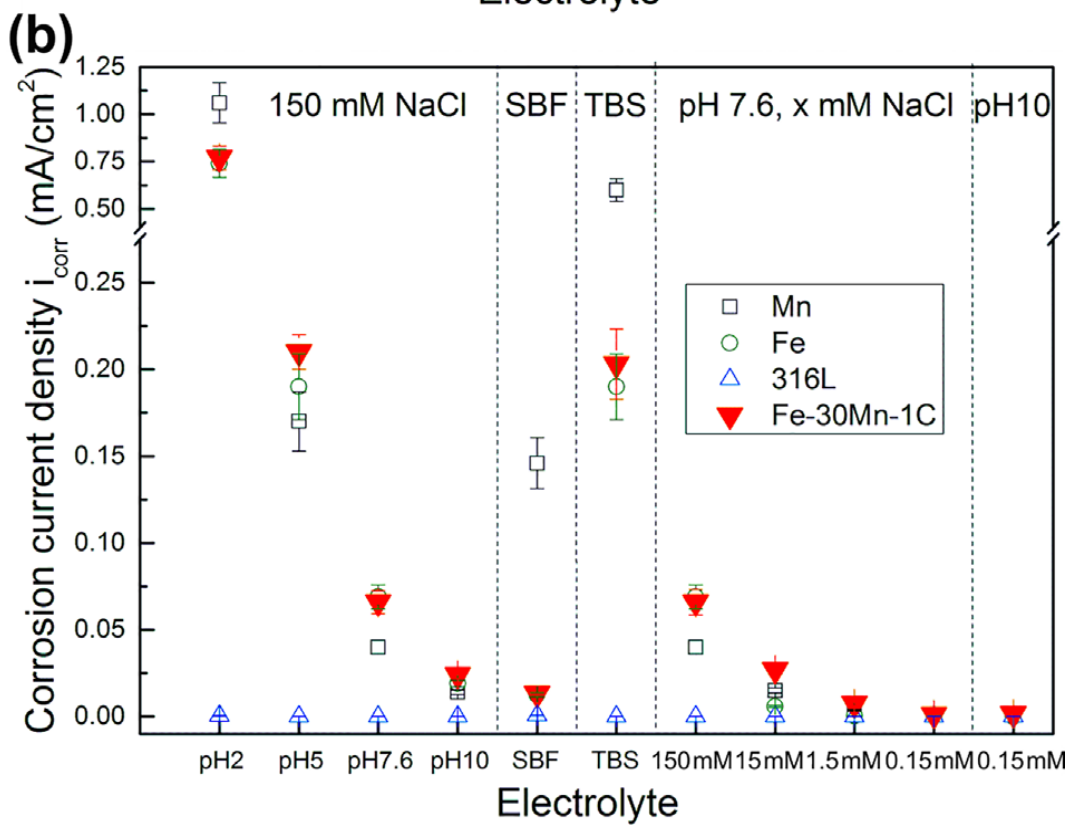

Figure 9

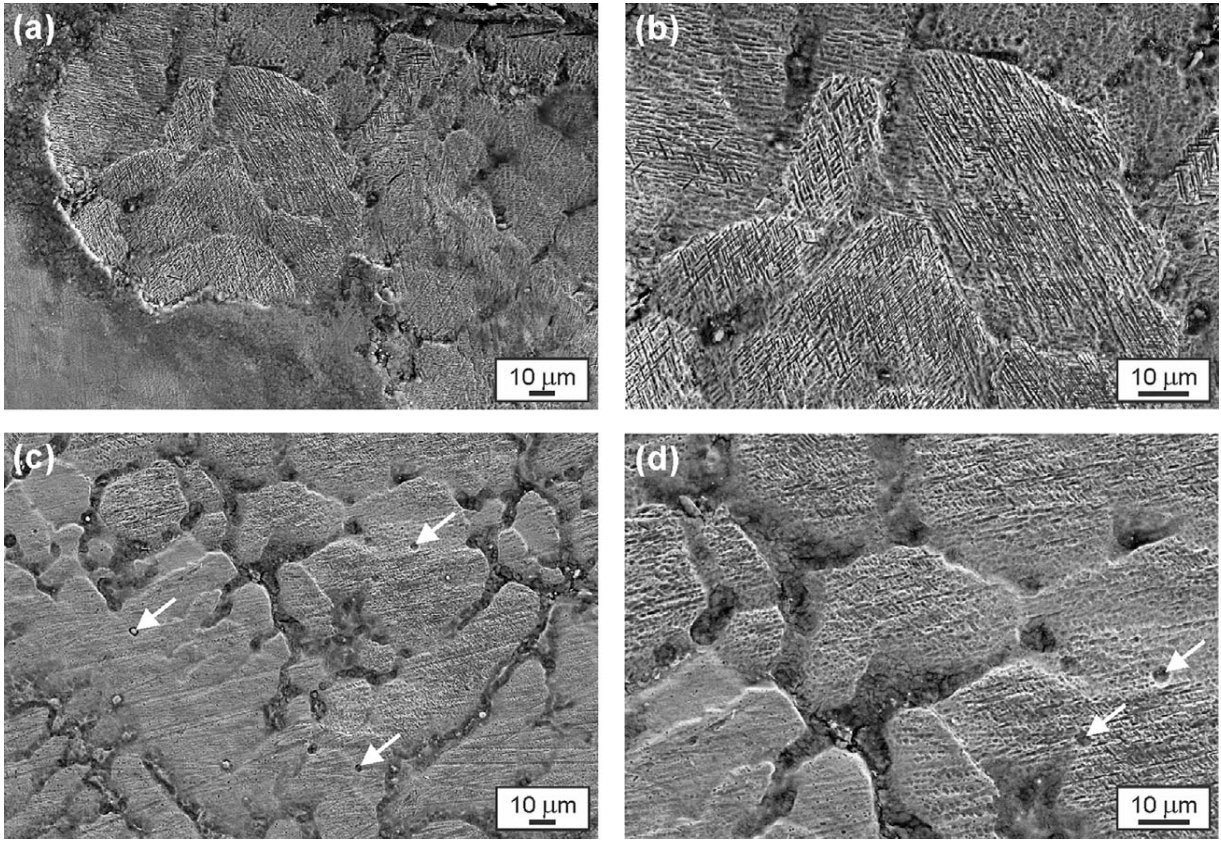

Figure 10